# Imaging through a thin scattering layer and jointly retrieving the point-spread-function using phase-diversity

Tengfei Wu,[1,2] Jonathan Dong,[2] Xiaopeng Shao,[1,*] and Sylvain Gigan [2,*]

[1] *School of Physics and Optoelectronic Engineering, Xidian University, Shaanxi 710071, China*
[2] *Laboratoire Kastler Brossel, ENS-PSL Research University, CNRS, UPMC Sorbonne Universités, Collège de France, Paris 75005, France*
*Corresponding author: xpshao@xidian.edu.cn    sylvain.gigan@lkb.ens.fr*

**Recently introduced angular-memory-effect based techniques enable non-invasive imaging of objects hidden behind thin scattering layers. However, both the speckle-correlation and the bispectrum analysis are based on the statistical average of large amounts of speckle grains, which determines that they can hardly access the important information of the point-spread-function (PSF) of a highly scattering imaging system. Here, inspired by notions used in astronomy, we present a phase-diversity speckle imaging scheme, based on recording a sequence of intensity speckle patterns at various imaging planes, and experimentally demonstrate that in addition to being able to retrieve diffraction-limited image of hidden objects, phase-diversity can also simultaneously estimate the pupil function and the PSF of a highly scattering imaging system without any guide-star nor reference.**

The interaction between light and complex samples with inhomogeneous refractive index in many imaging scenarios induces light scattering, which is always seen as an obstacle for imaging objects hidden inside or behind such samples and makes direct observation impossible, instead, generates a complex speckle pattern [1]. In recent years, wavefront shaping techniques have emerged as a powerful tool for imaging hidden objects or focusing through highly scattering media by controlling the incident light [2-11]. However, these techniques are complex and lengthy, since they require a detector or an optical/acoustical probe in the plane of interests. A recent breakthrough reported by Bertolotti et al. [12] avoided the use of guide-stars and enabled non-invasive imaging through thin scattering layers by exploiting the inherent angular-correlations, known as "memory effect" in the scattered speckle patterns [13, 14]. The angular signal is the convolution between the object that is placed within the range determined by the angular-memory-effect and the system's point-spread-function (PSF), which is a highly complex speckle pattern, as generated by any light point source on the object. Since the autocorrelation of the PSF is close to a $\delta$ function, the Fourier-amplitude of object is retrieved from a large speckle pattern (i.e. sufficient speckle grains) by calculating its autocorrelation, and the lost phase information is recovered via an iterative phase-retrieval algorithm [15]. Katz et al. [16] put forward a single-shot approach of the aforementioned concept, inspired by astronomical techniques in which they regarded the scattering imaging system as an incoherent imaging system, and directly obtain the signal on a camera. Bispectrum analysis [17] also allows imaging from a single image in the same scenario, by exploiting the fact that the Fourier-phase of an object can be deterministically extracted from a large speckle pattern by relying on the property that the bispectrum of the system's PSF is real-valued [18]. Although the speckle-correlation and the bispectrum analysis can retrieve hidden objects, they cannot directly access the scatterer's exact phase distortion, i.e. the PSF, since both methods are based on the idea of statistical average and the influence of PSF is eliminated or ignored in the reconstruction processes. Alternatively, if a light point source is present, one can measure the intensity PSF of the highly scattering imaging system [19] and then perform image reconstruction by deconvolution. However, it is inconvenient and unpractical to introduce such a guide-star in most real applications, e.g. in biomedical imaging. Yet, the prospect of accessing not only the object, but the scattering layer's properties, and the exact PSF, would be highly beneficial for non-invasive imaging, for instance to image more complex objects.

In this Letter, we experimentally demonstrate a non-invasive imaging scheme based on angular-memory-effect. Inspired by the phase-diversity technique used in astronomy imaging [20], a sequence of speckle patterns from multiple planes are sequentially collected by a translational camera. From only a small region of such speckle patterns sequence, we can jointly retrieve the diffraction-limited image of incoherently illuminated objects, hidden behind a thin, but highly scattering layer, and estimate the local PSF of the highly scattering imaging system, without any reference nor additional experimental constraint on the object side. In addition, our method is straightforward to implement and to the best of our knowledge, is the first experimental demonstration that phase-diversity technique can be used in highly scattering cases, well beyond the aberration regime.

The principle of the experiments for phase-diversity speckle imaging and a numerical simulation are presented in Fig. 1. An object, hidden at a distance $u$ behind a highly scattering medium, is illuminated by a spatially incoherent and narrowband source. The scattered light is recorded by a camera, which is initially placed at a distance $v$ from the other side of the scattering medium. Since the object is located within the range determined by angular-memory-effect, each point on the object generates a nearly identical, but shifted, random speckle pattern on the camera. The camera image is a simple incoherent superposition of these shifted random speckle patterns, generated by all the points on the object. Therefore, it allows the system in Fig. 1(a) to be regarded as an incoherent imaging system with a shift-invariant PSF, i.e. the identical random speckle pattern, and the camera image is the convolution of the object and the PSF [16]. In order to jointly retrieve the image of the object and estimate the PSF of the system, we continuously change the position of the camera with a fixed interval $\delta$ and collect the camera image $I_n(x)$ at each position (Fig. 1(b)), which reads:

$$I_n(x) = O(x) * S_n(x) + w_n(x) \quad (1)$$

where $n = 0, 1, 2, ..., N-1$ and $n = 0$ means the initial position. $O(x)$ is the image of object, $S_n(x)$ denote the corresponding PSFs of the system when camera is placed at different positions and $w_n(x)$ is a noise term. Since the speckle is very large (up to millions of speckle grains) the full problem is numerically extremely challenging. We therefore apply a phase-diversity algorithm to various sub-regions of the speckle patterns, and retrieve simultaneously the diffraction-limited image (Fig. 1(c)) and the corresponding sub-PSF (Fig. 1(d)) for the selected sub-region. Detailed information of the method is given below and in Fig. 2.

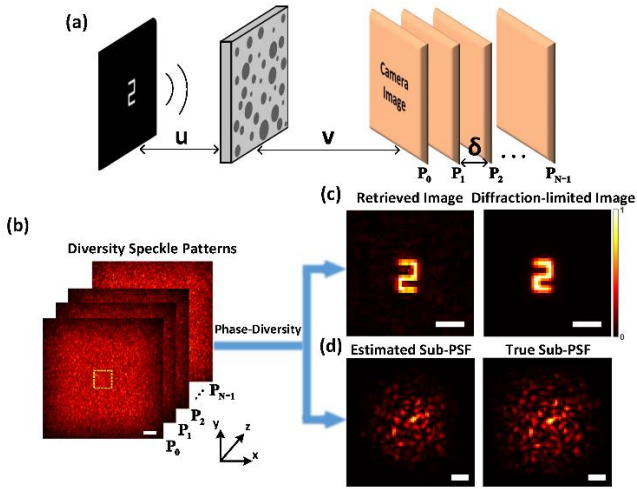

Fig. 1 Concept and numerical simulations: Jointly imaging hidden objects and estimating the PSF of a highly scattering imaging system. (a) An object is illuminated by a spatially incoherent light. The scattered light is recorded by a camera at various positions with a fixed interval $\delta$. P₀ is the initial position; (b) Simulated diversity speckle patterns, in which the selected sub-region is marked by a dashed box; (c) Retrieved image and diffraction-limited image; (d) Estimated sub-PSF and true sub-PSF (only the intensity is shown). Scale bar: 400 camera pixels in (b) and 10 camera pixels in (c) and (d).

As an early technique in astronomy imaging, phase-diversity has been used to estimate the unknown phase aberrations and eliminate the small distortions caused by the atmosphere turbulence, often by taking one or multiple snapshots around the focus, producing a set of blurred, but recognizable images for weak aberrations [20]. However, in highly scattering cases, only a 3D propagating speckle pattern is produced and sampled by the camera, containing no obvious visual information on the hidden object. Nonetheless, when introducing diversity, the generalized set of pupil functions of the highly scattering imaging system can still be written as follows in its most general form:

$$H_n(f) = |H_n(f)| \exp\{i[\phi(f) + \theta_n(f)]\} \quad (2)$$

where $f$ is the position in the pupil, $\phi(f)$ is the unknown random phases on the pupil, and $\theta_n(f)$ denotes the known phase function, corresponding to the $n$-th diversity image. We consider that the modulus of the generalized pupil function $|H_n(f)|$ is 1 over a round pupil and 0 elsewhere. The effective pupil diameter is determined by the lateral decorrelation length of speckle grains, $\Delta X = 1.0\lambda(v/D)$ [1], where $D$ denotes the diameter of the aperture stop placed between the scattering medium and the camera. In conventional phase-diversity (in the aberration regime), the unknown phases on the pupil of the system are smooth and can be conveniently parametrized with the basic functions of discretized Zernike polynomials. However, in our highly scattering case, the unknown phases $\phi(f)$ are totally random and the complex phase distribution is better sampled in the canonical (pixel) basis. In order to determine the local phase values, multiple diversity images are required to provide sufficient information. For simplicity and accessibility, we use a simple axial translation of the camera, which corresponds to the known parabolic diversity phase function of defocus [21]:

$$\theta_n(f) = \frac{n\delta\pi}{\lambda v(v+n\delta)} |f|^2 \quad (3)$$

where $\lambda$ is the wavelength of the incident light, $v$ is the imaging distance and $|f|$ is the distance to the center of the pupil.

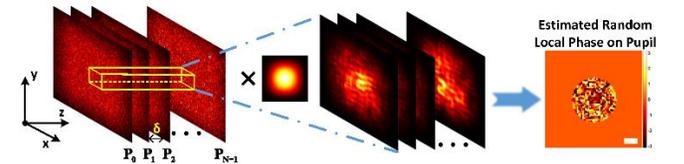

Fig. 2 Detailed information of the reconstruction of the speckle patterns in phase-diversity speckle imaging. Only a small region of speckle patterns is sufficient to reconstruct. Additional apodization function is required to smooth the edges of each sub-image. Scale bar: 10 camera pixels.

If we assume that the additive white Gaussian noise is the dominant noise (e.g. thermal noise) in the speckle patterns on the camera, a cost function can be found to estimate the local phase values on the pupil of the system via a maximum-likelihood estimation [20, 22]:

$$L[\phi(f)] = \sum_f \left[ \frac{\left|\sum_{k=0}^{N-1} \tilde{I}_k(f)\tilde{S}_k^*(f)\right|^2}{\sum_{j=0}^{N-1}|\tilde{S}_j(f)|^2 + \sigma} - \sum_{n=0}^{N-1} |\tilde{I}_n(f)|^2 \right] \quad (4)$$

where $\tilde{I}_{k,j,n}(f)$, $\tilde{S}_{k,j,n}(f)$ are the Fourier transforms of $I_n(x)$ and $S_n(x)$ in Eq. (1). "$*$" denotes the complex conjugation and $\sigma$ is a positive parameter to improve the convergence and stability of the optimization process. The PSF of the system is the squared modulus of the coherent impulse response function, which is the inverse Fourier transform of the generalized pupil function:

$$S_n(x) = |\mathcal{F}^{-1}\{H_n(f)\}|^2 \quad (5)$$

where $\mathcal{F}^{-1}\{\cdot\}$ implies the inverse Fourier transform. The image of the hidden object is simultaneously retrieved from [20, 22]:

$$O(x) = \mathcal{F}^{-1}\left\{\frac{\sum_{n=0}^{N-1}\tilde{I}_n(f)\tilde{S}_n^*(f)}{\sum_{k=0}^{N-1}\left|\tilde{S}_k(f)\right|^2+\sigma}\right\} \quad (6)$$

Interestingly, since the phase-diversity speckle imaging method is not based on the statistical average of speckle grains, only a small region of speckle pattern, as is marked by the dashed box in Fig. 2(a), is sufficient to retrieve the hidden object and simultaneously estimate a sub-PSF of the system, provided it is larger than the object autocorrelation. Larger sub-image can certainly provide more accurate estimation of PSF, yet simultaneously, increase the computational complexity. Considering a sub-region of the speckle pattern gives rise to an inevitable problem of the discontinuities of the edges of the sub-image [23]. The discontinuities produce artifacts in the Fourier transforms, which are implemented as two-dimensional, discrete fast Fourier transforms. These artifacts affect the accuracy of the reconstruction of the object and the estimation of the sub-PSF. The way in our method to overcome this problem is to smooth the edges of the sub-image by adding an apodization function (e.g. Hanning window). It is worth nothing that the lateral diffusion of scattered light during axial propagation would cause a boundary effect. More specifically, when moving the camera in the axial direction, the scattered light propagates and therefore diffuses laterally (together with a global translation for off-axis regions). This means that because we keep the size of phase-diversity images, there will always be a mismatch between the propagated images and the captured sub-images. In order to minimize the influence of the boundary effect, the selection of the sub-image is advantageously done near the optical axis and the maximum translation of camera from the initial position should be limited. It depends on the number of speckle grains contained in the sub-image and on the amount of translation. As an example, in our numerical simulation of Fig. 1, the lateral decorrelation length is 2.25 pixels, and there are 70 pixels in one dimension of sub-images, corresponding to around 31 speckle grains. 19 diversity images are collected corresponding in total to 6 axial decorrelation length, meaning in our case that we sample 6 speckle grains along the axial direction. This number is well below the number of speckle grains contained in one lateral dimension of the sub-image and the apodization window, ensuring a limited boundary effect.

In the experimental demonstration of this concept, the light source is a spatially incoherent light-emitting diode (Thorlabs, M625L3), whose nominal wavelength is 625nm and bandwidth is 18nm, filtered by a narrow band-pass filter (Andover, 633FS02-50, 1.0+/-0.2nm) mounted on the camera (Andor, ZYLA-5.5-USB 3.0), to ensure the high contrast of the speckle patterns. The incoherent light illuminates the object, digit "2" (Edmund, 1951 USAF Negative Target, 2" × 2", ~350um), which is shown in Fig. 4(a) and hidden ~60cm behind a scattering medium (Edmund, Ground Glass Diffuser). The illuminated area on the scattering medium is adjusted by a contiguous iris with a diameter of 5.21mm to control the size of speckle grains. The camera is initially placed at a distance of ~12cm in front of the scattering medium to collect the scattered light and a 50mm linear translation stage (Thorlabs, DDSM50) is used to move the camera.

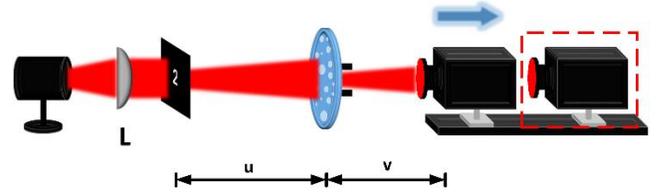

Fig. 3 Experimental set-up. A light-emitting diode is used as the spatially incoherent source, and a narrow band-pass filter is mounted on the camera to ensure the high contrast of the speckle patterns. The camera is moved step-by-step with a linear translation stage. L: Lens.

Fig. 4(b) shows some of the 19 acquired frames of raw diversity camera images with a fixed interval $\delta = \Delta Z/3$, where $\Delta Z = 7.1\lambda(v/D)^2$ is the axial decorrelation length [1]. The camera images are spatially normalized for the slowly varying envelope of the scattered light pattern, and then smoothed by a Gaussian kernel. We select three independent groups of sub-images around the center of the processed camera images. Each sub-image contains 84 camera pixels in one dimension and is smoothed by a Hanning window to solve the problem of discontinuities of edges in discrete Fourier transforms and limit the boundary effect. A quasi-Newton algorithm from Matlab Optimization Toolbox is used to solve the optimization problem in Eq. (4) to reconstruct each group of sub-images, respectively. The initial guess of the local phase values are zero and the parameter $\sigma$ is $10^{-5}$, which is selected by trial and error. If $\sigma$ is too small, the reconstructions degrade because of the amplification of error; while if it is too large, it overly smooths the reconstructions. The optimized random phases are estimated after 1200 iterations, as is shown in the first column of Fig. 4(c). The estimated sub-PSFs of the system and the retrieved images of hidden object are naturally obtained via Eq. (5) and Eq. (6).

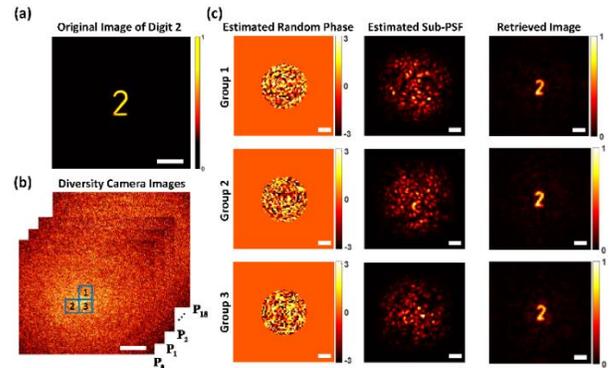

Fig. 4 Experimental results of phase-diversity speckle imaging. (a) Original image of hidden object, digit "2"; (b) Raw diversity camera images, in which three independent groups of sub-images are selected and reconstructed, respectively; (c) First column: estimated random local phase values, corresponding to different groups; second column: estimated sub-PSFs; third column: retrieved images of the hidden object from the three groups of speckle patterns. Scale bar: 500 camera pixels in (a) and (b); 10 camera pixels in (c).

As presented, our phase-diversity speckle imaging method can not only image the hidden object behind a thin, but complex scattering layer, as the speckle-correlation and the bispectrum analysis, but also estimate the pupil function and the PSF of a highly scattering imaging system without any reference. As long as the PSF is acquired, we can directly and efficiently recover the image of other hidden objects from the complex speckle pattern via simple deconvolutions, instead of repeating foregoing procedures or using other complex imaging methods.

As a demonstration of this concept, after estimating the PSF, we replace digit "2" with another object, digit "3" (~350um), which is from the same USAF target and shown in Fig. 5(a). The linear translation stage is used to move the camera back to the initial position to capture the speckle pattern (Fig. 5(b)). After implementing the same pre-processing to the raw camera image, we select the sub-images at the same areas, where the sub-PSFs have been estimated. The Hanning window is used to smooth each sub-image, as is shown in the first column of Fig. 5(c). Then, deconvolutions are implemented by applying the Lucy-Richardson method with the iterations between 30 and 50, using codes from the Matlab Image Processing Toolbox. Deconvolution results of the three groups are shown in Fig. 5(c), which demonstrates the feasibility of the deconvolution method, and moreover, verifies the validity of the estimated sub-PSFs via our phase-diversity speckle imaging method. Beyond this simple demonstration, more complex objects, and even broadband or polychromatic objects, could be in principle retrieved, since the local pupil phase function is known.

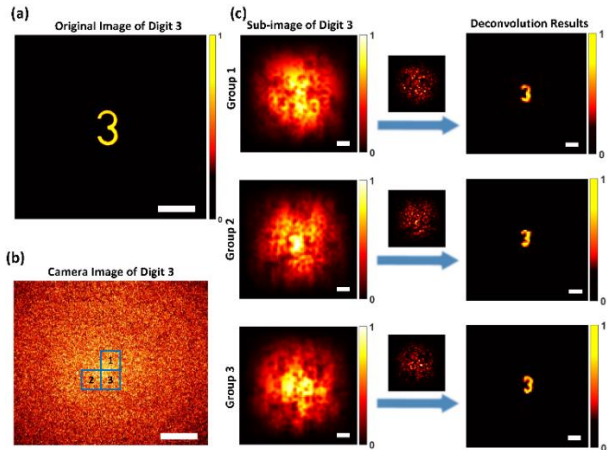

Fig. 5 Experimental results of imaging hidden object by simple deconvolution with the estimated sub-PSFs from phase-diversity speckle imaging method. (a) Original image of hidden object, digit "3", placed at the same position as digit "2"; (b) Raw camera image. Selecting the same three areas, where the sub-PSFs are estimated in advance; (c) First column: selected three sub-images, smoothed by the Hanning window; second column: retrieved results from three sub-images by deconvolution. Middle images are the corresponding sub-PSFs estimated by phase-diversity speckle imaging method as in Fig.4. Scale bar: 500 camera pixels in (a) and (b); 10 camera pixels in (c).

In conclusion, we experimentally demonstrated a phase-diversity speckle imaging method, which allows non-invasive imaging of hidden objects behind a thin, but highly complex scattering layer, and can jointly estimate the pupil function and the PSF of a highly scattering imaging system without any reference. Since our method is not based on the statistical average of large amounts of speckle grains, only a small region of speckle pattern is sufficient, at the cost of multiple frame acquisition. The smallest dimension of the sub-image depends on the image of hidden objects, whose size can be roughly determined by the autocorrelation of the camera image, and on the amount of translation performed. Larger region certainly allows us to acquire more accurate information of PSF, but on the other hand, it means more complex phase values on the pupil to determine, which increases the computational complexity. Furthermore, while we only obtain the local PSF within a sub-region of the speckle pattern, the process can easily be parallelized on multiple sub-apertures and a complete and high resolution reconstruction of the scatterer's topography could in principle be achieved, as in ptychography. For this proof of concept, we used a simple quasi-Newton algorithm to solve the optimization problem, which is possible to be trapped in the local minima and fail to converge. Slightly changing the available information (e.g. the frames of speckle patterns) or using random initial guesses would be potential solutions. In addition, other advanced optimization algorithms are expected to perform more efficiently. To this end, we freely make available the source codes and experimental data to use by the scientific community, as shown in the Codes and Data file.

**Funding.** National Natural Science Foundation of China (NSFC) (61575154); European Research Council (ERC) (278025 and 724473); China Scholarship Council (CSC) (201506960026);

**Acknowledgment.** The authors would like to thank Xueen Wang for the insightful discussions, and Huijuan Li for help with experiments. S. G. is a member of the Institut Universitaire de France.

REFERENCES

1. J. W. Goodman, Speckle phenomena in optics : Theory and applications (Roberts & Co., 2007).
2. I. M. Vellekoop, and A. P. Mosk, Opt Lett 32, 2309-2311 (2007).
3. I. M. Vellekoop, and C. M. Aegerter, Opt Lett 35, 1245-1247 (2010).
4. S. M. Popoff, G. Lerosey, R. Carminati, M. Fink, A. C. Boccara, and S. Gigan, Phys Rev Lett 104 (2010).
5. A. P. Mosk, A. Lagendijk, G. Lerosey, and M. Fink, Nat Photonics 6, 283-292 (2012).
6. O. Katz, E. Small, and Y. Silberberg, Nat Photonics 6, 549-553 (2012).
7. T. Chaigne, J. Gateau, O. Katz, E. Bossy, and S. Gigan, Opt Lett 39, 2664-2667 (2014).
8. P. X. Lai, L. D. Wang, J. W. Tay, and L. H. V. Wang, Nat Photonics 9, 126-132 (2015).
9. R. Horstmeyer, H. W. Ruan, and C. H. Yang, Nat Photonics 9, 563-571 (2015).
10. S. Rotter, and S. Gigan, Rev Mod Phys 89 (2017).
11. O. Katz, E. Small, Y. F. Guan, and Y. Silberberg, Optica 1, 170-174 (2014).
12. J. Bertolotti, E. G. van Putten, C. Blum, A. Lagendijk, W. L. Vos, and A. P. Mosk, Nature 491, 232-234 (2012).
13. S. Feng, C. Kane, P. A. Lee, and A. D. Stone, Phys Rev Lett 61, 834-837 (1988).
14. I. I. Freund, M. Rosenbluh, and S. Feng, Phys Rev Lett 61, 2328-2331 (1988).
15. J. R. Fienup, Appl Opt 21, 2758-2769 (1982).
16. O. Katz, P. Heidmann, M. Fink, and S. Gigan, Nat Photonics 8, 784-790 (2014).
17. T. Wu, O. Katz, X. Shao, and S. Gigan, Opt Lett 41, 5003-5006 (2016).
18. A. W. Lohmann, G. Weigelt, and B. Wirnitzer, Appl Opt 22, 4028 (1983).
19. E. Edrei, and G. Scarcelli, Sci Rep-Uk 6 (2016).
20. R. G. Paxman, T. J. Schulz, and J. R. Fienup, JOSA A 9, 1072-1085 (1992).
21. J. W. Goodman, Introduction to fourier optics (Roberts and Company Publishers, 2005).
22. C. R. Vogel, T. Chan, and R. Plemmons, P Soc Photo-Opt Ins 3353, 994-1005 (1998).
23. O. Von der Lühe, Astronomy and Astrophysics 268, 374-390 (1993).

Informational fifth page

REFERENCES
1. J. W. Goodman, Speckle phenomena in optics : Theory and applications (Roberts & Co., 2007).
2. I. M. Vellekoop, and A. P. Mosk, "Focusing coherent light through opaque strongly scattering media," Opt Lett 32, 2309-2311 (2007).
3. I. M. Vellekoop, and C. M. Aegerter, "Scattered light fluorescence microscopy: imaging through turbid layers," Opt Lett 35, 1245-1247 (2010).
4. S. M. Popoff, G. Lerosey, R. Carminati, M. Fink, A. C. Boccara, and S. Gigan, "Measuring the Transmission Matrix in Optics: An Approach to the Study and Control of Light Propagation in Disordered Media," Phys Rev Lett 104 (2010).
5. A. P. Mosk, A. Lagendijk, G. Lerosey, and M. Fink, "Controlling waves in space and time for imaging and focusing in complex media," Nat Photonics 6, 283-292 (2012).
6. O. Katz, E. Small, and Y. Silberberg, "Looking around corners and through thin turbid layers in real time with scattered incoherent light," Nat Photonics 6, 549-553 (2012).
7. T. Chaigne, J. Gateau, O. Katz, E. Bossy, and S. Gigan, "Light focusing and two-dimensional imaging through scattering media using the photoacoustic transmission matrix with an ultrasound array," Opt Lett 39, 2664-2667 (2014).
8. P. X. Lai, L. D. Wang, J. W. Tay, and L. H. V. Wang, "Time-reversed ultrasonically encoded optical focusing into scattering media," Nat Photonics 9, 126-132 (2015)
9. R. Horstmeyer, H. W. Ruan, and C. H. Yang, "Guidestar-assisted wavefront-shaping methods for focusing light into biological tissue," Nat Photonics 9, 563-571 (2015).
10. S. Rotter, and S. Gigan, "Light fields in complex media: Mesoscopic scattering meets wave control," Rev Mod Phys 89 (2017).
11. O. Katz, E. Small, Y. F. Guan, and Y. Silberberg, "Noninvasive nonlinear focusing and imaging through strongly scattering turbid layers," Optica 1, 170-174 (2014).
12. J. Bertolotti, E. G. van Putten, C. Blum, A. Lagendijk, W. L. Vos, and A. P. Mosk, "Non-invasive imaging through opaque scattering layers," Nature 491, 232-234 (2012).
13. S. C. Feng, C. Kane, P. A. Lee, and A. D. Stone, "Correlations and Fluctuations of Coherent Wave Transmission through Disordered Media," Phys Rev Lett 61, 834-837 (1988).
14. I. Freund, M. Rosenbluh, and S. Feng, "Memory Effects in Propagation of Optical Waves through Disordered Media," Phys Rev Lett 61, 2328-2331 (1988).
15. J. R. Fienup, "Phase Retrieval Algorithms - a Comparison," Appl Optics 21, 2758-2769 (1982).
16. O. Katz, P. Heidmann, M. Fink, and S. Gigan, "Non-invasive single-shot imaging through scattering layers and around corners via speckle correlations," Nat Photonics 8, 784-790 (2014).
17. T. Wu, O. Katz, X. Shao, and S. Gigan, "Single-shot diffraction-limited imaging through scattering layers via bispectrum analysis," Opt Lett 41, 5003-5006 (2016).
18. A. W. Lohmann, G. Weigelt, and B. Wirnitzer, "Speckle Masking in Astronomy - Triple Correlation Theory and Applications," Appl Optics 22, 4028-4037 (1983).
19. E. Edrei, and G. Scarcelli, "Memory-effect based deconvolution microscopy for super-resolution imaging through scattering media," Sci Rep-Uk 6 (2016).
20. R. G. Paxman, T. J. Schulz, and J. R. Fienup, "Joint estimation of object and aberrations by using phase diversity," JOSA A 9, 1072-1085 (1992).
21. J. W. Goodman, Introduction to fourier optics (Roberts and Company Publishers, 2005).
22. C. R. Vogel, T. Chan, and R. Plemmons, "Fast algorithms for phase diversity-based blind deconvolution," P Soc Photo-Opt Ins 3353, 994-1005 (1998).
23. O. Von der Lühe, "Speckle imaging of solar small scale structure I. Methods," Astronomy and Astrophysics 268, 374-390 (1993).